\documentstyle[12pt, epsf]{article}

\begin{document}


\centerline{\bf Thermal metastabilities in the solar core}
\vskip 1cm
\centerline{Attila Grandpierre}
\centerline{grandp@konkoly.hu}
\centerline{\it Konkoly Observatory of the Hungarian Academy of Sciences}
\centerline{ H-1525 Budapest, P. O. Box 67, Hungary}
\vskip 1cm
\centerline{Gabor Agoston}
\centerline{\it Eastron TDA}
\centerline{H-1148 Budapest, Nagy Lajos Kir\'aly u. 20, Hungary}
\date{\today}
\vskip 2cm
\centerline{\large Abstract} 
Linear stability analysis indicates that solar core is thermally stable for infinitesimal internal perturbations. For the first time, thermal metastabilities are found in the solar core when outer perturbations with significant amplitude are present. The obtained results show that hot bubbles generated by outer perturbations may travel a significant distance in the body of the Sun. These deep-origin hot bubbles have mass, energy, and chemical composition that may be related to solar flares. The results obtained may have remarkable relations to activity cycles in planets like Jupiter and also in extrasolar planetary systems.
\vskip 2cm

\newpage
"The Sun's interior is believed to be in a quiescent state and therefore the relevant physics is simple. However, after decades of work, we are still not certain whether the physics of solar energy generation is as simple as expected" - wrote Bahcall \cite{Bahcall89}. The linear analysis \cite{Schwarzschild73, Rosenbluth73, Paterno97}  indicated that solar core is stable to internal radial and non-radial perturbations. At the same time, the Sun is thought to be unstable for a wide range of magnetic and rotational instabilities and changing in a wide variety of timescales \cite{Gough90}. Above a slightly larger than actual central temperature of the Sun, at $1.74 \times 10^7 K$ CNO would begin to dominate over pp cycles. One may expect that outer perturbations generating local heating may set up convection-like movements in the solar core. 
Grandpierre \cite{Grandpierre90} suggested that planetary influences may interact with local magnetic fields in the solar core, and the arising local heating may generate temperatures above $10^8$ K. The possible role of solar inertial motions (SIM) around the mass centre of the solar system in the solar activity was recently emphasised \cite{Zaqarashvili97, Charvatova00, Juckett00}. In a frame fixed to the Sun the waves generated by SIM have a relative velocity $\approx 1.5 \times 10^5$ cm/s, which is the rotational velocity of the solar surface. Their dissipated energies are derived from the rotational energy of the Sun. Solar rotation during the last $4.6 \times 10^9$ years has been spun down from the earlier 10 times the present rotation rate with $E_{rot,0}  \approx 10^{45}$ ergs, representing $L_{rot,0} \approx 2 \times 10^{-6}L_{Sun}$. 
Although the present spun-down Sun has $L_{rot} \simeq 2 \times 10^{-8} L_{Sun}$, it is still possible that at "elementary dissipation events" localised in space and time local rotation energy dissipations may become significant at the most favourite sites where magnetic and planetary perturbation inhomogenities are the most pronounced. Considering bubbles with $V_b > 10^{18} cm^3$ (i.e. the minimal bubble volume to reach solar surface, $V_c$  is the volume of the solar core $\approx 10^{31}$ $cm^3$), $V_c/V_b < 10^{13}$, the dissipation luminosity has an upper limit $L_{diss} < L_{rot} \times 10^{13} \approx 10^5$ $L_{Sun}$. Moreover, the episodic character of dissipation events offer a concentration in time in a rate $C_t \approx 1$ to $10^6$ leading to $L_{diss} < 10^{11}$ $L_{Sun}$. For a significant local heating to occur, we can allow to go down many orders of magnitude from the above derived upper limit.  For example, with $L_{diss} \approx 10^{-1} L_{Sun}$, it is possible to reach a bubble initial temperature $T_0 > 1.7 \times 10^7$ K. 

The motion of a heated bubble is determined by the equality of the buoyant ($F_b=Vg \Delta \rho$) and frictional ($F_f=K/2 v^2 S \rho_s$) forces, where $S$ is the cross section of the bubble, $V$ is its volume, $K$ is the coefficient of turbulent viscosity, $\rho_s$ and $\Delta \rho$  are the density of the surroundings and the density difference between the bubble and its surroundings, and $g$ is the gravitational acceleration. Equating these forces, $v^2 K/2 S/V = g \Delta \rho / \rho_s$. Now assuming pressure equilibrium between the bubble (referred with no index) and its surroundings (referred with index s), $\rho T = \rho_s T_s$, $\Delta \rho/\rho_s = (1-T_s/T)$. Taking $K = 1$, we obtain for the bubble's velocity $v=(8/3Rg(1-T/T_s))^{1/2}$. With typical values in the solar core $g= 2\times 10^5 cm s^{-2}$, $R=10^5-10^6$ cm,  $v \approx 2$ to $7 \times 10^5$ $cm s^{-1}$ \cite{Gorbatsky64}. 
The estimation of thermal adjustment time is given by the following formula of \cite{Kippenhahn90}:
\begin{eqnarray}
\tau_{adj} = C_p \kappa \rho^2 R^2/(4acT^3).
\end{eqnarray}
With typical values ($\kappa = 2 cm^2 g^{-1}, \rho = 90 g cm^{-3}, T = 10^8 K, 
R = 10^6$ cm), $\tau_{adj} = 3 \times 10^3$ s, while for $T = 10^7$ K, 
$\tau_{adj} = 3 \times 10^6$ s.
We may define a surfacing time-scale for the bubbles
\begin{eqnarray}
\tau_{surf} = l_T/v,
\end{eqnarray}
where $l_T$ is the temperature scale height in the solar core $l_T \approx  1.5 \times 10^{10}$ cm. With $v=1.5 \times 10^5 cm s^{-1}$, $\tau_{surf} \approx 10^5 $ s. 

Estimating the nuclear heating time-scale of a hot bubble from $\epsilon =\epsilon_0 \rho T^{\nu}=C_p \partial T/\partial t$, with a substitution $x=T^{\nu}$, $(1/x) \partial x/\partial t = \epsilon \nu/C_p T$, we obtained a relation \cite{Grandpierre90} to the local time-scale of the nuclear heating of the hot bubble as \begin{eqnarray}
\tau_{nucl}=C_p T/ \epsilon \nu.
\end{eqnarray}
When T changes from $10^7$ K to $10^9$ K, $\tau_{nucl}$ changes from $10^{16}$ s to $10^{-1}$ s. 
We calculated the adiabatic expansion of the hot bubbles following Gorbatsky \cite{Gorbatsky64}, and the related time-scale is
\begin{eqnarray}
\tau_{ad} = -5/(1/p dp/dr) 1/(8/3(1-T_s/T)Rg)^{1/2}
\end{eqnarray}
$\tau_{ad}$ is usually in the range of $10^4$ to $10^5$ s. From these estimations we can see that there are four relevant time-scales and these may be comparable for thermal perturbations in the solar core. Therefore, they indicate the possibility of thermal metastability and the need for detailed numerical calculations.

We regarded that the material heated by the heat wave of radiative diffusion expanding from the bubble is coupled to it \cite{Gorbatsky64}. We worked with a fourth order Runge-Kutta method to solve the differential equation system. They are based on:
\begin{eqnarray}
dR/dt = (-1/5) (1/p dp/dr) v R = R/\tau_{adj}
\end{eqnarray}

\begin{eqnarray}
dQ/dt = (dQ/dt)_{ad} + (dQ/dt)_{eps} + \epsilon m
\end{eqnarray}

\begin{eqnarray}
dQ/dt \approx 2/5 Q/p dp/dr + 3/2 R_g/\mu T_S 4 \pi \rho_S R^3/\tau_{adj} + \epsilon 4\pi/3 R^3 \rho
\end{eqnarray}

\begin{eqnarray}
v=(8/3 Rg(1-T_s/T))^{1/2}
\end{eqnarray}

We used the opacity-figure of \cite{Rogers98} and approximated it as $\kappa = 10^{10.45}/T^{1.45}$ above $3 \times 10^5$ K. Our calculations differ from the previous ones that we considered perturbations for which $\rho' \not= 0$, and $T' \not= 0$. This is a difference to the case considered elsewhere \cite{Schwarzschild73, Rosenbluth73, Paterno97} with $\rho'= 0$, and $(T/\mu)'=0$. In our calculations, we allowed the transient development of pressure inequilibrium. The duration of this pressure inequilibrium is a small fraction of a second, and therefore practically we have pressure equilibrium, $p'=0$, $(T \rho/\mu)' = 0$, but $\rho' \not= 0$ and $T'\not= 0$.  
We solved the differential equation system with a numerical code developed by one of us (G. A.). We neglected the radiation pressure in all the terms except the diffusive one. Our method works well below and around $10^8$ K, the estimated errors in each quantities are smaller than $15 \%$. At higher temperatures, the approach applied becomes more uncertain and therefore we could not follow the development of the hot bubbles at thermal runaways where our method indicated temperatures above $10^{10}$ K. 

We may observe in TABLE I. that above $10^8$ K the heating time-scale becomes lower than that of the combined (expansion plus diffusion) cooling one. Our detailed calculations has shown that:

1.) Hot bubbles formed in the solar core may travel significant distances. Already with $T_0 = 1.74 \times 10^7$ K the bubbles take a path of $\approx 1.2 \times 10^{10}$ cm, i.e. even a moderate local heating may trigger hot bubbles travelling far from their site of birth. Depending on the size, energy and frequency distribution of initial perturbations, the solar core may become dynamic (meta-convection).

2.) Hot bubbles may reach the solar surface even with $Q_0=10^{35}$ ergs. The internal energy excess of the bubbles when arriving to subphotospheric regions may reach $10^{32}$ - $10^{36}$ ergs, where the bubble's velocity is typically around $5$ - $15$ $km s^{-1}$. 

3.) For initial temperatures higher than $10^8$ K numerical calculations indicate the presence of thermonuclear instability. 

4.) Hot bubbles have a tendency to preserve their chemical composition corresponding to deeper layers, and consume a part of their hydrogen content, and therefore they will have "anomalous" chemical composition.

Our results demonstrate the possibility of a coupling between some high velocity hot bubbles and some activity centres like rigidly rotating \cite{Spence93} sunspot nests \cite{Castenmiller86, De Toma00}, hot spots \cite{Bai95}, active longitudes \cite{Bai95}, and perhaps even with some individual atmospheric eruptions like flares, coronal mass ejections of protuberances.  The possibility of such relations raises the question of relation between mass, energy and chemical composition of the bubbles and such atmospheric events. The initial mass of a hot bubble is around $10^{18}$ - $10^{20}$ gram. Therefore, if its mass is conserved during its path to the envelope, in order to find a connection to flares with mass $10^{15}$ - $10^{16}$ g, a mechanism should exist between the suphotospheric and coronal regions to concentrate energy into a tiny fraction of mass. One candidate for such a mechanism is offered by the interesting fact that the bubbles approach the speed of sound when entering the subphotospheric regions. The developing sonic boom generates shock waves, transforming the energy of high speed bubbles into energy of shock waves and energetic particle beams \cite{Grandpierre91}.

The bubble's cores when approaching the surface have a size of $\approx 10^7$ cm at $r \approx 0.93 R_{Sun}$. In the subphotospheric regions they grow to $\approx 10^8$ cm, in good agreement with the photospheric sizes of flare kernels $\approx 2 000$ km. Besides the energetic, mass and size relations, some connections between the solar neutrino problems, as well as the chemical anomalies of many flares and that of the hot bubbles may also be related. 
One may speculate that SIM inertial tides play an important role in the generation of solar activity. The periodic acceleration and deceleration of SIM may drive an inertial oscillation in the solar core. Tidal stresses are maximal at the solar center \cite{Peale78}. Magnetic fields may act as a restoring force, therefore magnetic torsional oscillation may arise which may explain the solar cycle \cite{Gough90}. The maximum value of the Sun's orbital angular momentum reaches 25$\%$ of the Sun's rotational momentum $1.7 \times 10^{47} g cm^2 s^{-1}$. Spin-orbit coupling, a transfer of angular momentum from the Sun's orbit to the Sun's spin rate and vice versa could make a difference of more than $5 \%$ in the equatorial velocity which is actually observed \cite{Landscheidt88, Gilman84, Hathaway90}. Interestingly, hot spots occur at the Earth as well, they also show deep origin from the core \cite{Morgan71}, rigid rotation \cite{Jurdy84}, chemical anomalies \cite{Hoffmann97} and surface activity. A significant positive correlation exists between the chemically 'anomalous' activity of Jupiter's hot spots and the solar cycle \cite{Kostiuk00}. The phenomenon of active longitudes is known also in close binaries where the position of the two diametrically opposing active longitudes is directly related to the position of the companion star (Olah, K. et al., in preparation). There is evidence for a purely tidal effect as well-enhanced activity at the subbinary point of close binaries \cite{Cuntz00}. Tidal response calculations are also of interest in connection with the newly discovered planets \cite{Terquem98}, which may play a role in the generation of the activity cycle of their host stars.

\eject

\begin{table}
\caption{The combined (expansion plus diffusion) cooling time-scale $\tau_{cool}$  in seconds for hot bubbles with different size and initial temperatures. For comparison, the heating (nuclear) time-scale $\tau_{nucl}$  is indicated at each temperature below the relevant cooling times in the second column. $T_s=1.318 \times 10^7$ K is the temperature of the surroundings at $r=0.1R_{Sun}$ from where the bubble starts rising in our calculations. Letter s indicates that the bubble reaches the solar surface, and letter r indicates the case when even thermal runaway occurs.   }
\vskip 1cm
\begin{tabular}{lllllll}
$T_0$[K] &$\tau_{nucl}$[s] & & & $\tau_{cool}$ [s] & &\\	
\hline
& & $R_0=10^6$cm &$2*10^6$cm  &$3*10^6$cm&	$4*10^6$cm&	$5*10^6$cm\\
\hline
$1.0001T_s$ &$1.4 \times 10^{16}$		&$6.3 \times 10^6$	&$4.5 \times 10^6$	&$3.7 \times 10^6$	&$3.2 \times 10^6$ 	&$2.8 \times10^6$\\

$5 \times 10^7$		&$6.9 \times 10^7$		&$1.9 \times 10^4$	&$2.9 \times 10^4$	&$2.9 \times 10^4$	&$2.7 \times 10^4$	&$2.4 \times 10^4$\\

$10^8$ 		&$1.1 \times 10^5$		&$3.2 \times 10^3$	&$1.0 \times 10^4$	&$1.5 \times 10^4$	&$1.8 \times 10^4$s	&$1.8 \times 10^4$s\\

$2 \times 10^8$ 		&$7.8 \times 10^1$		&$4.2 \times 10^2$	&$1.6 \times 10^3$	&$3.4 \times 10^3$s	&$5.4 \times 10^3$s	&$7.2 \times 10^3$r\\

$3 \times 10^8$		&2.6			&$1.3 \times 10^2$	&$5.0 \times 10^2$r	&r &r &r\\

$4 \times 10^8$ 		&$6.5 \times 10^{-1}$		&$5.3 \times 10^1$s	&$2.1 \times 10^2$r	&r &r &r\\

$5 \times 10^8$		&$3.4 \times 10^{-1}$		&$2.7 \times 10^1$r	&r		&r		&r		&r\\

\end{tabular}
\end{table}
	
\begin{figure}
\epsfxsize15cm
\epsfbox{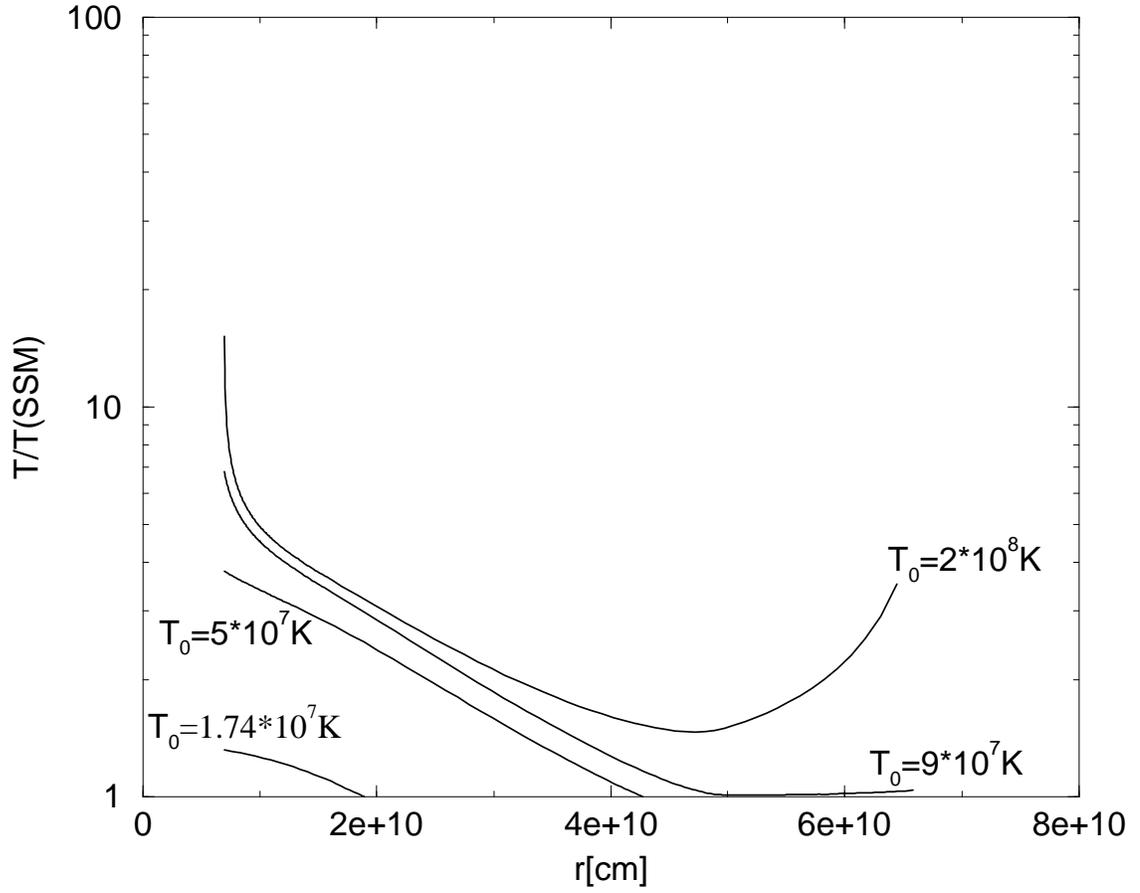}
\caption{ The run of the relative temperature of the hot bubbles for $T_0=1.74 \times 10^7 K$, $5 \times 10^7 K$, $9 \times 10^7 K$ and $2 \times 10^8 K$, when $R_0=3 \times 10^8 cm$. T and T(SSM) are the temperatures of the hot bubbles, and their standard solar model surroundings, respectively, r is the distance from the solar center.}
\end{figure}

\begin{figure}
\epsfxsize15cm
\epsfbox{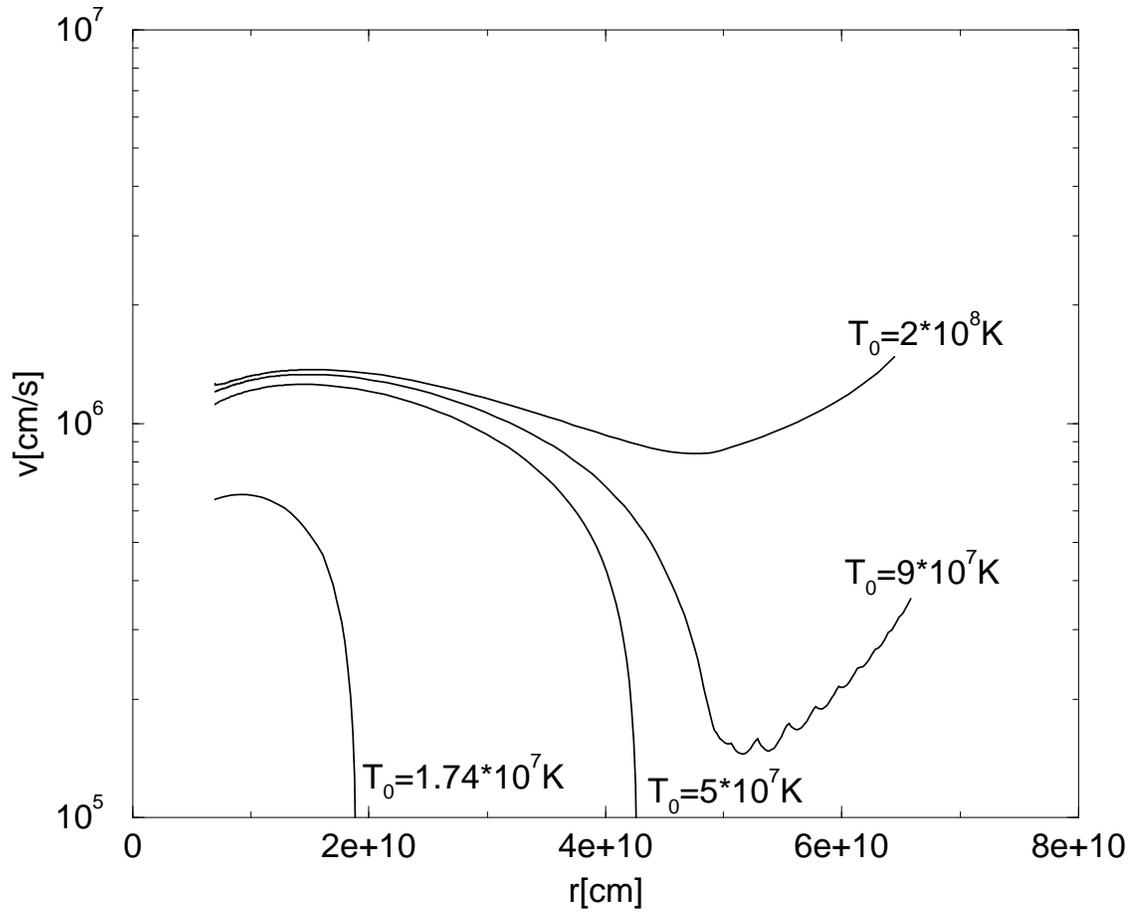}
\caption{ The run of the hot bubbles' rising speed. The small waves around the end of our curves do not represent a phbysical effect as they are due to the linear interpolation applied in the solar model.}
\end{figure}

\begin{figure}
\epsfxsize15cm
\epsfbox{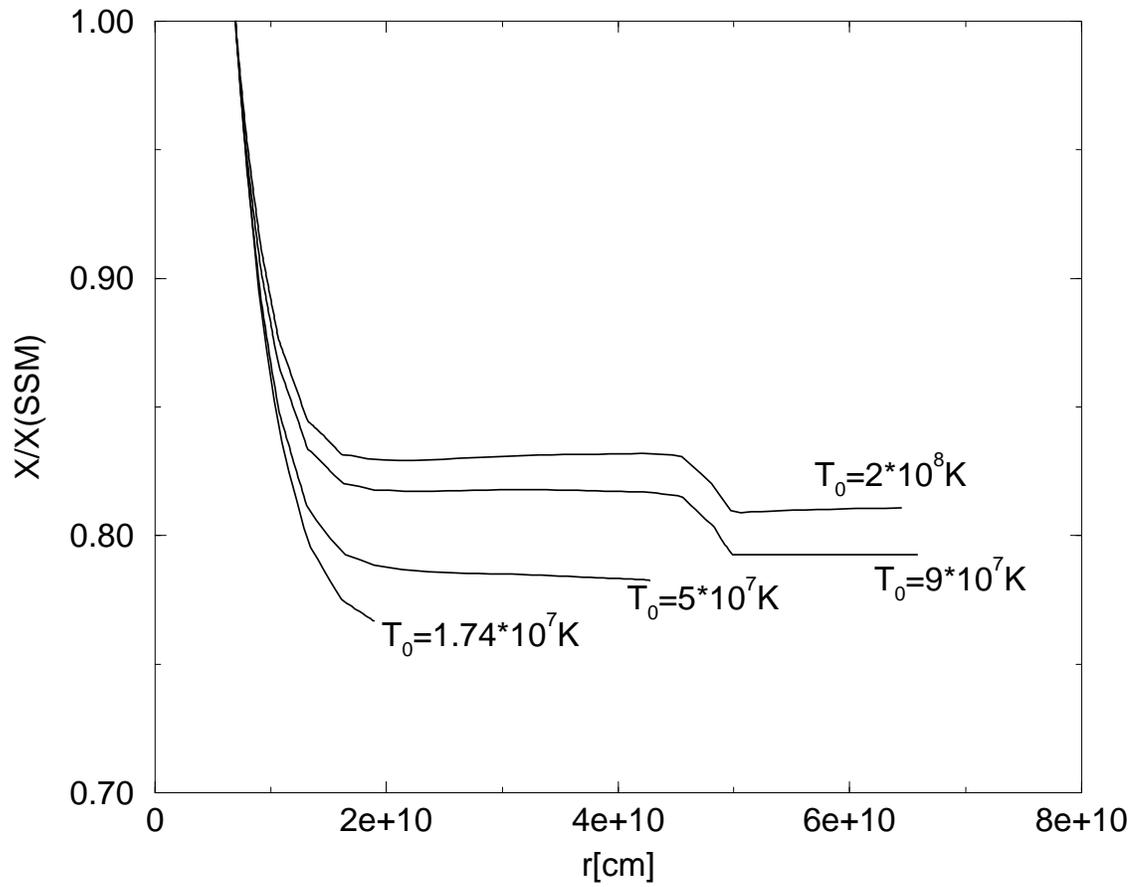}
\caption{ The run of the relative hydrogen abundances X/X(SSM)  of the hot bubbles. }
\end{figure}

\eject

\end{document}